\documentclass[showpacs,12pt,preprintnumbers,amsmath,amssumb,floatfix]{revtex4}
\usepackage{graphicx} 
\usepackage{dcolumn}
\usepackage{bm}

\begin{document}

\title{Symmetry-broken crystal structure of elemental boron at low temperature}

\author{M. Widom}
\affiliation{Department of Physics, Carnegie Mellon University, Pittsburgh,
PA, USA}
\author{M. Mihalkovi\v{c}}
\affiliation{Institute of Physics, Slovak Academy of Sciences, Bratislava,
Slovakia}

\date{\today}

\begin{abstract}
The crystal structure of boron is unique among chemical elements, highly
complex, and imperfectly known.  Experimentalists report the
$\beta$-rhombohedral (black) form is stable over all temperatures from
absolute zero to melting.  However, early calculations found its
energy to be greater than the energy of the $\alpha$-rhombohedral
(red) form, implying $\beta$ cannot be stable at low
temperatures. Furthermore, $\beta$ exhibits partially occupied sites,
seemingly in conflict with the thermodynamic requirement that entropy
vanish at low temperature.  Using electronic density functional theory
methods and an extensive search of the configuration space we find a
unique, energy minimizing pattern of occupied and vacant sites that
can be stable at low temperatures but that breaks the
$\beta$-rhombohedral symmetry.  Even lower energies occur within
larger unit cells. Alternative configurations lie nearby in energy,
allowing the entropy of partial occupancy to stabilize the
$\beta$-rhombohedral structure through a phase transition at moderate
temperature.
\end{abstract}

\maketitle

\section{Introduction}
Elemental boron is important for its light weight, high strength, high
melting point and semiconducting properties.  It is also intrinsically
interesting owing to its complex structures which are characterized by
their arrangements of icosahedral clusters.  Boron is almost unique
among elements in lacking a well understood and universally agreed
upon low temperature structure.  Knowledge of the precise structure is
required for understanding its remarkable electronic and mechanical
properties, and also for understanding the stability of
technologically important boron compounds relative to their
constituent elements.

\begin{figure}
\includegraphics[width=7in]{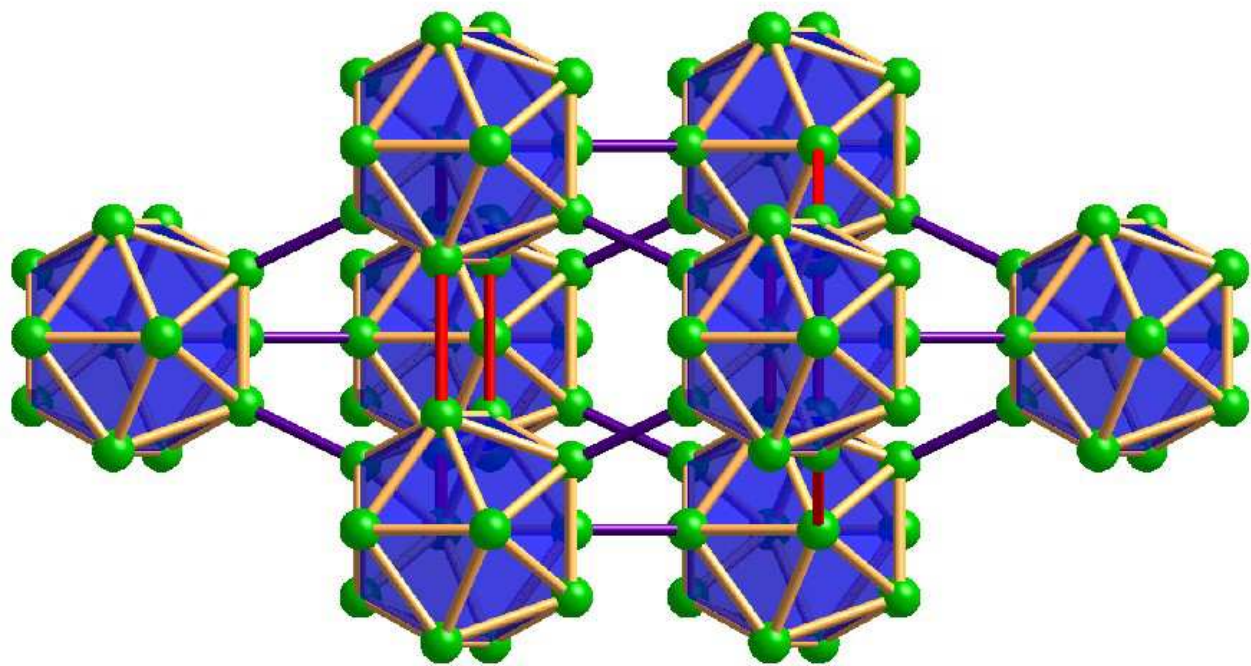}
\caption{\label{fig:alpha} Structure of $\alpha$-rhombohedral boron
viewed along the rhombohedral (11$\bar{2}$) axis. Bond color scheme:
1.67~\AA~ in purple; 1.75-1.81~\AA~ in orange; 2.01~\AA~ in red.}
\end{figure}

The first reported crystallographic refinements of elemental boron
were the $\alpha$-tetragonal~\cite{Hoard51,Hoard58} and
$\alpha$-rhombohedral forms~\cite{McCarty,Decker}.  Although the
$\alpha$-rhombohedral form was initially believed stable at low
temperature~\cite{McCarty}, the $\beta$-rhombohedral form, was later
discovered~\cite{Hughes} and proposed as the true low temperature
state.  Still later, the $\beta$-tetragonal form was discovered and
its crystal structure refined~\cite{Vlasse}.  The $\alpha$ and
$\beta$-rhombohedral structures are illustrated in
Figs.~\ref{fig:alpha} and~\ref{fig:beta} and are described in greater
detail in Section~\ref{sec:structure}.

The debate over the proper stable form of boron continues today, with
some researchers suggesting $\alpha$-rhombohedral is the true low
temperature state~\cite{Masago06}.  In particular, the calculated
energy of $\alpha$ lies below the reported calculated energies of
$\beta$~\cite{Masago06,Mihalkovic04,Jemmis05,Setten07}.  Some researchers
propose a finite temperature phase transition~\cite{Werheit86},
including one proposal that vibrational entropy drives an $\alpha$ to
$\beta$ transition at finite temperature~\cite{Masago06}.  Quantum
mechanical zero point vibrational energy has been proposed as a
mechanism to stabilize the $\beta$-rhombohedral form at absolute
zero~\cite{Setten07}.

We note that the $\beta$-rhombohedral form exhibits intrinsic disorder
in the form of partially occupied sites.  Partial occupancy occurs in
crystallographic refinement when the site is occupied, at a given
instant, in some copies of the unit cell but not in others.  Within a
single unit cell the site may be occupied at some times but not at
others. Since thermodynamics requires that entropy vanish in the limit
of low temperature, and partial occupancy implies finite entropy, the
partially occupied $\beta$-rhombohedral form is not a plausible low
temperature structure.  Correlations among the partially occupied
sites of $\beta$ must favor a unique pattern of occupancy that
minimizes the energy.

We carry out an extensive study of the configuration space that
explores specific resolutions of partial occupancy, assigning atoms or
vacancies to specific sites.  Our results indicate that a particular
symmetry-broken form of $\beta$ achieves energy lower than $\alpha$
and thus is the true low temperature structure.  A symmetry restoring
phase transition, driven by the high entropy of an ensemble of nearly
degenerate configurations, should occur at a moderate temperature,
explaining the experimental observation of $\beta$-rhombohedral as the
equilibrium phase at higher temperatures.

The remainder of this introduction briefly describes the known
structures of crystalline boron and our calculational methods.
Results are presented in Section~\ref{sec:results}.  We find optimal
assignemnts of atoms to partially occupied sites, we validate these
findings using solid state molecular dynamics simluations, and we
explore possible superlatttice ordering.  Finally we discuss our
conclusions in Section~\ref{sec:discuss}.

\subsection{Structural description}
\label{sec:structure}

Crystalline boron occurs primarily in tetragonal and rhombohedral
forms.  Since the tetragonal forms are known to be metastable, our
analysis in this paper focuses on the rhombohedral forms.  Both the
$\alpha$ and $\beta$-rhombohedral forms share a common space group,
$R\bar{3}m$ (group \# 166).  Three rhombohedral primitive cells
combine to form a larger nonprimitive hexagonal unit cell with three
times the number of atomic sites.  The rhombohedral (111) axis becomes
the hexagonal symmetry axis of the hexagonal unit cell.  For
simplicity our analysis is based on the rhombohedral primitive cell,
though we do at the end consider supercells of the primtive cell
including the full hexagonal unit cell.

Structure types are denoted by their Pearson symbols. For example,
Pearson type tP50 ($\alpha$-tetragonal) is primitive tetragonal with
50 atomic sites per cell.  Pearson tP196 ($\beta$-tetragonal) is
also primitive tetragonal but with 196 atomic sites per unit cell.
Because of partial occupancy tP196 actually has fewer than 196 atoms
distributed among its 196 sites.  Pearson hR12 ($\alpha$-rhombohedral)
and hR105 ($\beta$-rhombohedral) are both rhombohedral primitive cells
of 12 and 105 sites repsectively.  Three rhombohedral primitive cells
may be combined to form a single hexagonal cell.

All crystalline boron structures share a common structural motif, the
12-atom icosahedral cluster.  Various allotropes differ in the spatial
arrangement of the icosahedra and in the presence of interstitial
atoms.  The structural complexity of boron is due to the ``electron
deficiency'' of the B$_{12}$ icosahedral
cluster~\cite{Higgins55,Lipscomb81} which frustrates the distribution
of electrons among available bonds.  Presumably the partial site
occupancy serves to relieve this frustration~\cite{Jemmis01}.

Among boron allotropes, $\alpha$-rhombohedral boron has the simplest
structure (see Fig.~\ref{fig:alpha}), with a B$_{12}$ icosahedral
cluster placed at each vertex of the rhombohedral cell.  The structure
is defined by just two independent Wyckoff positions~\cite{Wyckoff}.
The icosahedral clusters are nearly regular, with bond lengths in the
range of 1.75-1.81~\AA.  Clusters are joined along six of their twelve
5-fold axes that point radially outwards through vertices of the
icosahedron.  These intericosahedral bonds have length 1.67~\AA.  Also
visible in the figure are bonds of length 2.01~\AA~ running parallel
to the icosahedral 2-fold axes.  This bond length is almost completely
absent in all other allotropes, so we believe it is energetically
unfavorable.

\begin{figure}
\includegraphics[width=7in]{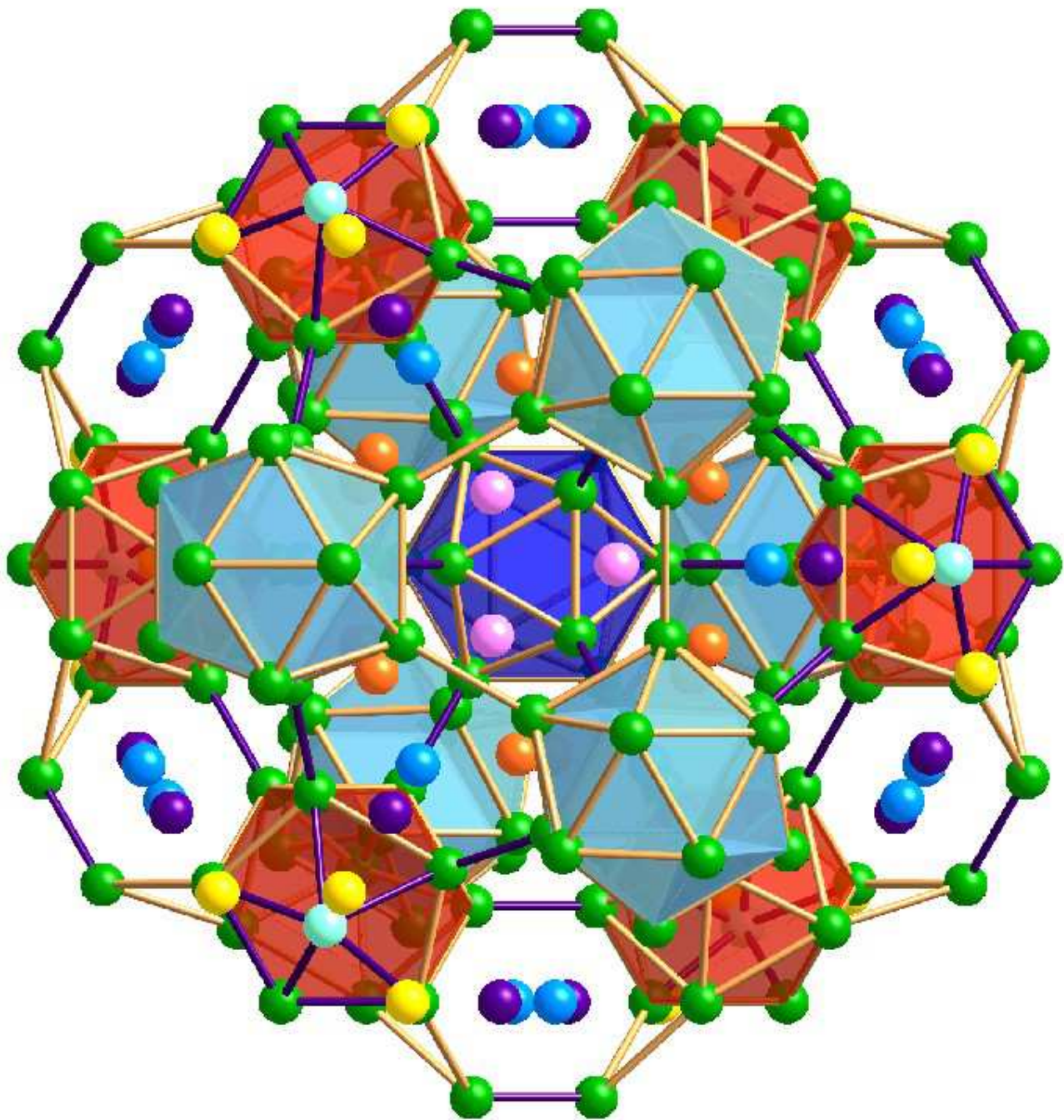}
\caption{\label{fig:beta} Structure of $\beta$-rhombohedral boron
viewed along rhombohedral (111) axis.  Bond color scheme:
1.63-1.73~\AA~ in purple; 1.73-1.92~\AA~ in orange.  Partially
occupied sites (see text) are shown in color: B13 (74.5\% average
occupancy) cyan; B16 (27.2\%) pink; B17 (8.5\%) yellow; B18 ( 6.6\%)
indigo; B19 (6.8\%) blue; B20 (3.7\%) orange. }
\end{figure}

The $\beta$-rhombohedral structure (see Fig.~\ref{fig:beta}) differs
from $\alpha$ in that the icosahedral cluster at the origin joins to
twelve other icosahedra along each of its twelve five-fold axes.  In
contrast to $\alpha$, where all icosahedra share a
common orientation, in $\beta$ the twelve surrounding icosahedra are
each rotated by 36$^{\circ}$ relative to the central icosahedron
around their common axes.  Consequently, the lattice constant of
$\beta$ is approximately double that of $\alpha$.  As shown, the
icosahedra shaded in blue are at positions equivalent to those in the
$\alpha$ structure, while those shaded in red are new.

\begin{figure}
\includegraphics[width=7in]{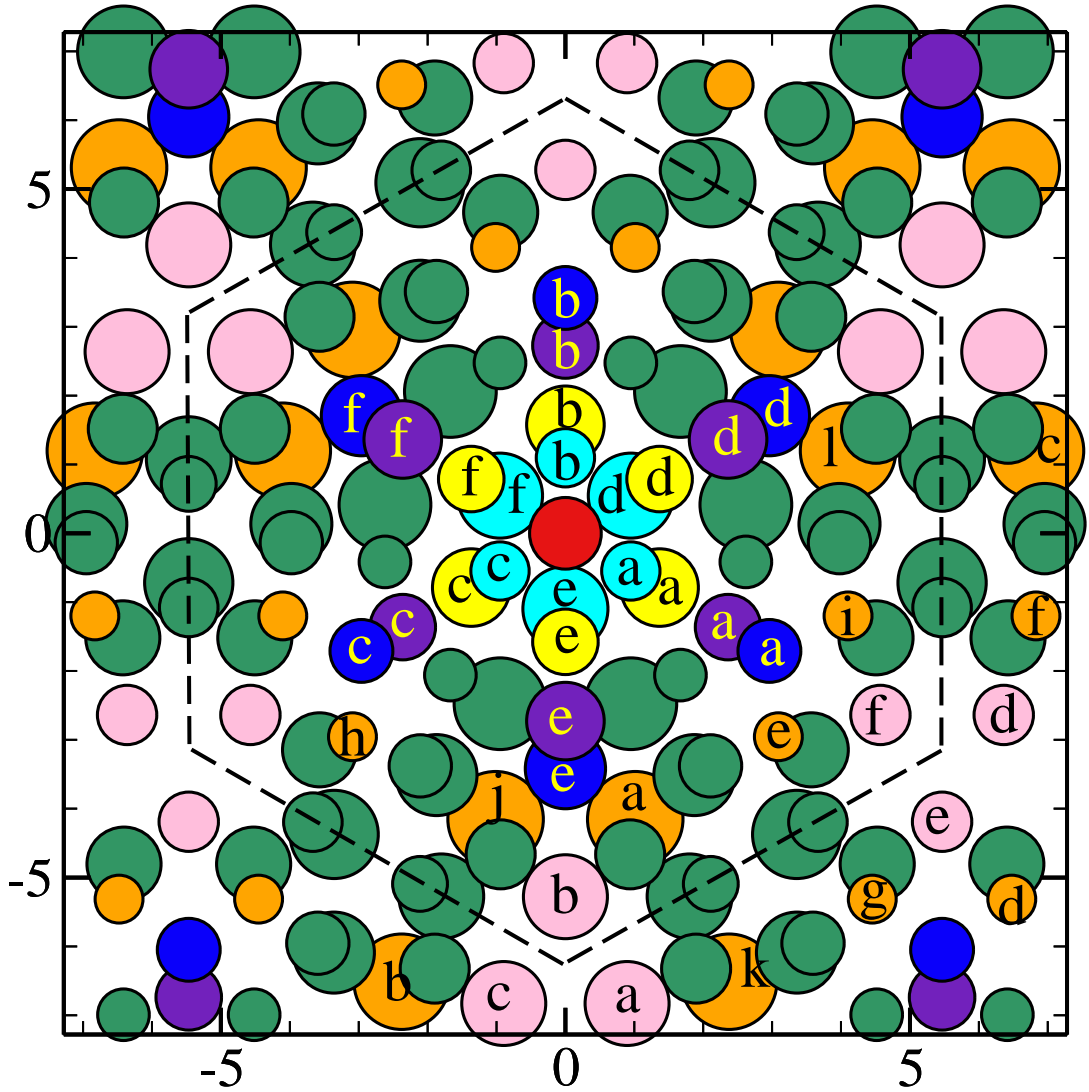}
\caption{\label{fig:B15} Structure of $\beta$-rhombohedral boron in
vicinity of cell body center (B15, shown in red) viewed along the
rhombohedral (111) axis.  Color coding as in Fig.~\ref{fig:beta}.
Site labels correspond to notation in table~\ref{tab:nrgs}.  Size of
atoms indicates vertical position, with small on top.  Length scale is
in Angstroms.  The dashed lines contains a single hexagonal unit cell.}
\end{figure}

Another new feature in $\beta$ is the presence of partially occupied
sites.  Although the original hR105 structure
model~\cite{Hughes,Geist} contained 15 Wyckoff positions, each fully
occupied, an intermediate model~\cite{Hoard70,Callmer} (Pearson type
hR111) assigned the B13 position 73\% average occupancy and introduced
a new B16 position at 25\% occupancy (the numbering scheme we use is
common to all the cited authors).  In the Pearson type hR141
model~\cite{Slack} illustrated in Fig.~\ref{fig:beta}, the B13
position is listed at 74.5\% average occupancy, the B16 position at
27.2\%, and additional positions B17, 18, 19 and 20 are given
occupancies of 8.5\%, 6.6\%, 6.8\% and 3.7\% respectively.  Other than
the B20 position, all partially occupied positions lie in mirror
planes of the structure that contain the rhombohedral (111) axis.
They appear as collinear sets in the projections shown in Figs.~\ref{fig:beta}
and~\ref{fig:B15}.

\subsection{Methods}
\label{sec:methods}

Our basic calculational methods follow Ref.~\cite{Mihalkovic04}.  The
calculations use the electronic density functional program
VASP~\cite{Kresse96} version 4.6.28.  VASP uses a plane-wave approach
that relies on periodic boundary conditions and is well suited to the
study of periodic crystal structures.  Electron-ion interactions are
represented using Projector Augmented Wave (PAW)
potentials~\cite{Blochl94,Kresse99} which are an all-electron
generalization of pseudopotentials.  For boron the 1s electrons are
treated within the ionic core while the 2s and 2p electrons are
assigned to the valence band.  All structures considered were
electrically neutral.

The exchange-correlation functional is taken as the PW91 Generalized
Gradient Approximation (GGA)~\cite{PW91}.  Previously the GGA has been
shown superior to LDA (the Local Density Approximation) for studies of
Boron clusters and compounds~\cite{Boustani95,Marlid01}, with accuracy
nearly that of Hartree-Fock calculations at the self-consistent-field
(SCF) level~\cite{Boustani97}.

All structures are fully relaxed, both in atomic coordinates and
lattice parameters, subject to the preservation of initial symmetry,
using a conjugate gradient algorithm.  Owing to coupling of the basis
set to the volume we perform consecutive calculations (i.e. stop and
restart) to fully relax the structure.  Reported energies are obtained
from a final static calculation.

We test the dependence of energy {\em differences} among the $\alpha$,
$\beta$-rhombohedral and optimized symmetry-broken $\beta$ (see
section~\ref{sec:hR141}) structures on computational parameters.  The
results are summarized in table~\ref{tab:conv}.  To maintain
consistent cell sizes we use a 2x2x2 supercell of the $\alpha$ phase,
whose lattice parameters become similar to those of $\beta$.  The
plane wave energy cutoff $E_{cut}$ is 319 eV by default, and we also
test 415 eV.  ``Precision'' settings of ``Medium'' and ``Accurate''
set the density of FFT grids which control wrap-around errors.  We
systematically increase the Monkhorst-Pack $k$-point mesh starting
from 1x1x1 (i.e. the $\Gamma$-point) until sufficient convergence is
achieved.  Based on the data presented we claim energy differences are
converged to within 6\% (roughly two significant figures) provided we
use a $k$-point mesh of 3x3x3 and ``Accurate'' precision.  The default
plane wave energy cutoff of 319 eV suffices.

In Table~\ref{tab:pot} we test the dependence on choice of potential
and exchange-correlation functional.  The local density
approximation~\cite{CA80} (LDA) is expected to be the least accurate
method.  The ultrasoft pseudopotential method~\cite{Vanderbilt90}
(USPP) treats only the valence electrons explicitly.  The ``hard''
potential has a very small core radius and correspondingly high
$E_{cut}$=700eV.  Results with this potential are expected to be
comparable in accuracy to all-electron FLAPW calculations and to
Gaussian with large basis sets, according to VASP
documentation~\cite{VASPweb}.

Because our final energy difference $E_{opt}-E_{\alpha}$ is not much
larger than the variation among the GGA calculations it would be
desirable to repeat this calculation using more refined
quantum-chemical methods.  However, all methods agree that our
optimized structure achieves a lower energy than $\alpha$, with the
exception of LDA which is expected to be the least accurate.  Even if
a more accurate calculation were to find that {\em all} variants of
$\beta$ were higher in energy than $\alpha$, that would not alter our
central conclusion that the fully symmetric $\beta$-rhombohedral
structure is a high temperature phase stabilized by occupancy
fluctuations.

\section{Results}
\label{sec:results}

Total energy calculations depend on precise knowledge of atomic
positions, with partially occupied sites resolved into a specific
pattern of occupied or vacant positions.  Likewise partial site
occupancy is thermodynamically forbidden in the $T=0K$ limit.  Hence
we explore the ensemble of likely instantaneous configurations,
seeking both the unique optimal arrangement of atoms among partially
occupied sites as well as an estimate of the entropy associated with
nearly optimal configurations.  We carry out the study initially
within a single rhombohedral primitive cell, repeated infinitely owing
to the periodic boundary conditiions.  Later we study superlattice
ordering within supercells.

Table~\ref{tab:nrgs} lists our main results~\cite{www}.  Energies are
given relative to the $\alpha$-rhombohedral form, which we take as a
reference because it contains no partial occupancy and was previously
the lowest energy structure that was known.  As expected, the
tetragonal structures, known to be metastable, exhibit relatively high
energies.  The fully occupied hR105 $\beta$-rhombohedral structure is
higher in energy than the $\alpha$-rhombohedral hR12 structure, as
previously noted~\cite{Masago06,Mihalkovic04,Jemmis05,Setten07}.  If
$\beta$ is to be stable at low temperatures it must involve the
placement of atoms among the partially occupied sites of hR111 or
hR141.  Some particular pattern of occupied and vacant sites must
minimize the total energy while breaking the rhombohedral symmetry.

The following tables present energy data supporting specific
conclusions on the optimization of B.hR111 and B.hR141.  Figures and
atomic coordinates for each named structure, and many other structures
not listed here, can be viewed at our web site~\cite{www} (see the
special ``published'' area).  Data is shown for symmetry-inequivalent
structures.  All energies are given for a 3x3x3 Monkhorst-Pack
$k$-point mesh, for accurate precision and the standard PAW potential
with default plane wave energy cutoff of 319 eV as discussed in
Section~\ref{sec:methods}.

The notation lists only those sites among the partially occupied
Wyckoff positions that are actually occupied.  Positions B1-B12, B14
and B15 are always fully occupied and thus are not listed in our
notation.  The hR111 structure introduces partial occupancy at B13 and
the new B16 position.  The hR141 structure introduces additional
positions B17-B20.  The notation $abcdef$ refers to specific sites
within each Wyckoff position, as labeled in Fig~\ref{fig:B15}.

\subsection{Pearson hR111}
\label{sec:hR111}
We first explore the hR111 model~\cite{Callmer}.  The
two partially occupied sites, B13 and B16 are both of Wyckoff type
18h, meaning that 18/3=6 of these sites occur per rhombohedral
primitive cell.  The B13 sites form a pair of equilateral triangles
surrounding the B15 site at the primitive cell body center (see light
blue atoms in Fig.~\ref{fig:B15}).  The B16 sites form a pair of
equilateral triangles that lie immediately above faces of the
icosahedra at primitive cell vertices (see pink atoms in
Fig.~\ref{fig:beta}).

A B13 atom can be swapped for a B16 atom in 4 symmetry-inequivalent
ways, each of which lowers the energy by 12-13 meV/atom.  This
substitution is thus strongly preferred energetically, but the spatial
correlation between occupied and vacated site is relatively weak, as
suggested by Slack~\cite{Slack}.

Since the number of atoms per cell is believed to be greater than 105,
we also considered a single B13 vacancy and a pair of B16 atoms, again
exhaustively testing all combinations.  Low energy requires that one
atom reside on each equilateral triangle of B16.  Our optimal
structure within the confines of hR111 is B13bcdefB16bd, at an energy
of 0.15 meV/atom above hR12.  The B13bcdefB16bf structure was the best
found in a previous study by van Setten, et al~\cite{Setten07}.

\subsection{Pearson hR141}
\label{sec:hR141}
So far the occupancy of B13 is larger than reported experimentally,
the energy remains above the energy of $\alpha$, and the total number
of atoms per primitive cell remains below the experimentally reported
value of 106.7.  Evidently we should try removing another atom from
B13 and placing that atom (and more) at other locations.  To place
these atoms we utilize the additional partially occupied positions of
hR141, namely B17-20.  There are of order $10^6$ distinct arrangements
of atoms within a single primitive cell that are consistent with
experimentally observed occupancies. Since this is far too many
structures to explore exhaustively, we build upon our prior results
and consider other likely correlations in order to focus our search.

Following Slack~\cite{Slack} we note the 1.57~\AA~ bond between B13
and B17 sites is slightly too short for simultaneous occupancy.  If we
choose to occupy B17a, this model suggests that B13d and B13e should
be vacant.  However, our calculations show this arrangement is not
stable.  One of the vacancies moves to the B13a site leaving just a
single vacancy on either B13d or e but not both.  Then the B17a atom
relaxes to accommodate the bond to the remaining nearby B13 atom,
which ends up at length 1.86~\AA.  It is noteworthy that the B17 site
reports an anomalously large thermal Debye-Waller factor, indicating
large displacements from the refined position.  We suggest that this
position should be split into distinct sites whose occupancy is
correlated with the nearby B13.  In fact, Slack utilized an
alternative split site ``B17d'' to refine one of his samples, and one
of these B17d sites lies within 0.1~\AA~ of our relaxed B17 position.

The 1.62~\AA~ bond length between sites B17 and B18 is within the
favorable range, and their reported occupancies are similar, so we
presume their occupation is correlated.  Our calculated energies
confirm this, with a reduction in energy of 1.1 meV/atom upon
introduction of an atom at the B18 site adjacent to an occupied B17
site.  After optimizing the placement of B16 atoms this leads to our
optimal primitive cell structure, B13bcefB16bdB17aB18a, which contains
107 atoms and achieves an energy 0.86 meV/atom below the energy of
$\alpha$.  This model also achieves a close match of relaxed positions
to experimentally observed positions, as is evident from the $\Delta
R$ values in Table~\ref{tab:nrgs}.

The B19 and B20 sites so far remain unused in our study.  B19 sites
are only 0.74~\AA~ away from B18 sites, so these may never be
simultaneously occupied.  Both B19 and B20 sites lie close to the
centers of fully occupied hexagonal rings, resulting in small patches
of triangular lattice that are atypical of crystalline boron
structures (although they may be stable in small boron clusters and
nanotubes~\cite{Boustani03}).  We found no significantly low energy
structures utilizing B19 or B20 sites.  Van Setten et
al.~\cite{Setten07} report a low energy for B19bcdefB16eB19a, but we
find this is not favorable. Rather the B19a atom relaxes to the
nearby B17a position.

\subsection{Molecular dynamics}

Because we cannot systematically evaluate all configurations within
the hR141 primitive cell, we checked our result using molecular
dynamics.  We used the VASP-calculated forces to perform molecular
dynamics simulations of a single primitive of $\beta$.  To achieve
atomic diffusion we employed a high temperature T=2000K (melting is
around T=2365K). We estimated the lattice parameters as
$a$=11.047~\AA~ and $c$=24.155~\AA~ based on an experimental report of
thermal expansion~\cite{Lundstrom98}.

We ran samples of 106, 107 and 108 atoms for a duration of 16ps each,
using 1fs time steps.  To search for optimal configurations we drew
instantaneous configurations from the molecular dynamics run every 2ps
and quenched them.  To carry out the quench we rescaled the lattice
parameters to their crystallographic values and performed molecular
dynamics with a linearly decreasing temperature ramp that reached
T=300K after 3ps.  We then performed conjugate gradient relaxation to
reach 0K.  This procedure was able, on occasion, to achieve the optimal
structures we reported, but usually resulted in higher energies.

\begin{figure}
\includegraphics[width=4in]{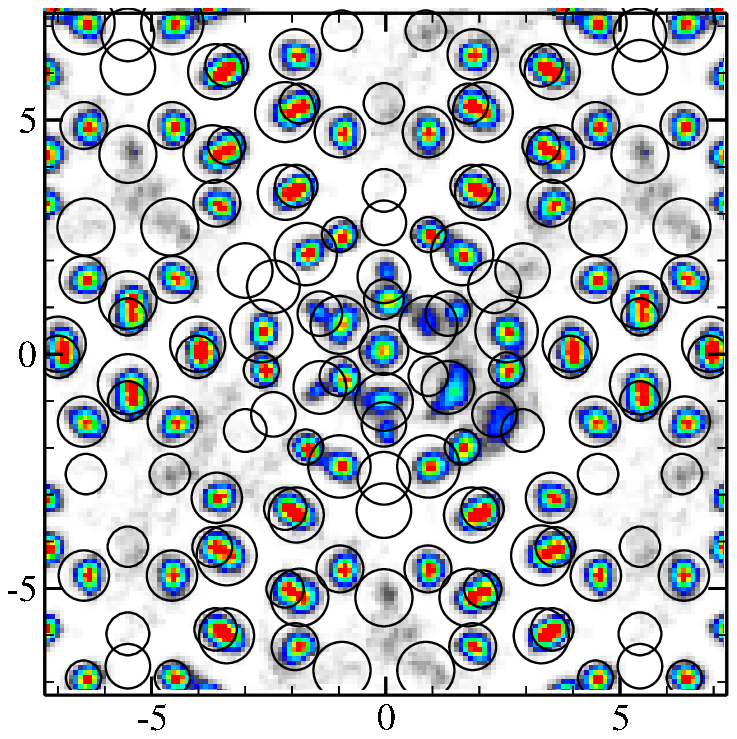}
\includegraphics[width=4in]{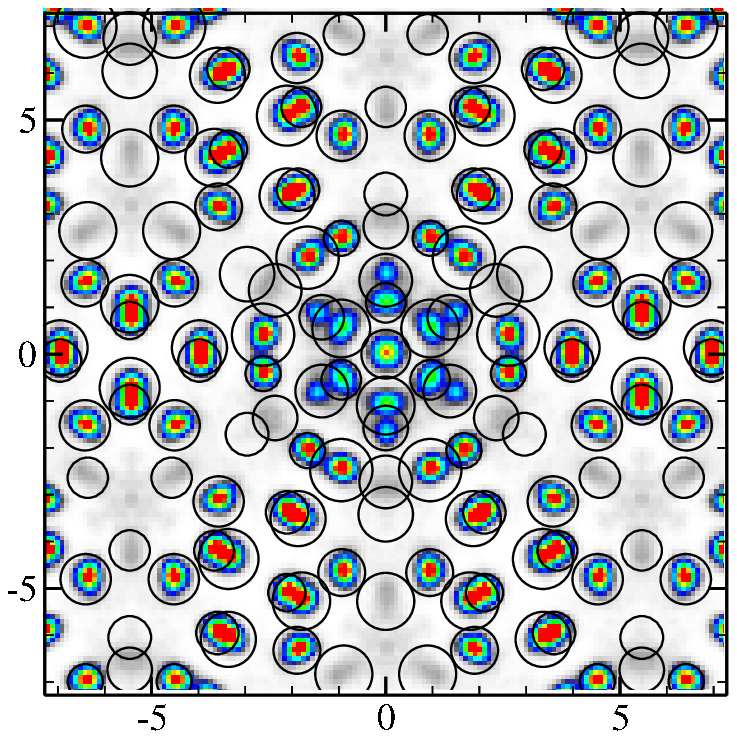}
\caption{\label{fig:md} Density plot of simulated atomic positions.
Grayscale indicates low frequency positions, colors indicate medium
(blue) - high (red) frequency positions.  Crystallographically
determined atomic positions are superimposed (and suitably scaled) for
comparison with Fig.~\ref{fig:B15}. (top) Run started in optimal
configuration. (bottom) Same data with rhombohedral symmetry imposed by
averaging.}
\end{figure}

Fig.~\ref{fig:md} shows a density plot of atomic positions for the 107
atom run that started with our optimal structure as an initial
condition.  Densities have been averaged to impose rhombohedral
symmetry.  Sharing of atoms between B13 and B17 positions is clearly
visible, as is partial occupation of B18 and B16 positions.

\subsection{Supercell studies}
\label{sec:scell}

Although we optimized the assignment of partial occupancy within a
single hR141 primitive cell, there is a possibility of a lower energy
structure within either the hexagonal unit cell (Pearson type hP423)
or some other supercell of the primitive cell.

The hexagonal unit cell (Pearson type hP423, dimensions $a$=10.9~\AA,
$c$=23.8~\AA) contains three copies of the rhombohedral primitive cell
stacked in the direction parallel to the rhombohedral (111) axis.  We
consider the case where each primitive cell is identically decorated,
where one is rotated by 120$^{\circ}$ relative to the other two, and
where each is rotated by 120$^{\circ}$ relative to its neighbors. In
this last case a $3_1$ screw axis is introduced.  In
Table~\ref{tab:ucell} energies are given for a 3x3x1 $k$-point mesh
with medium, maintaining a uniform reciprocal space density comparable
to a 3x3x3 mesh for a single primitive cell.  It appears energetically
preferable to maintain identical orientations of all vertically
stacked cells.

\begin{figure}
\includegraphics[width=7in]{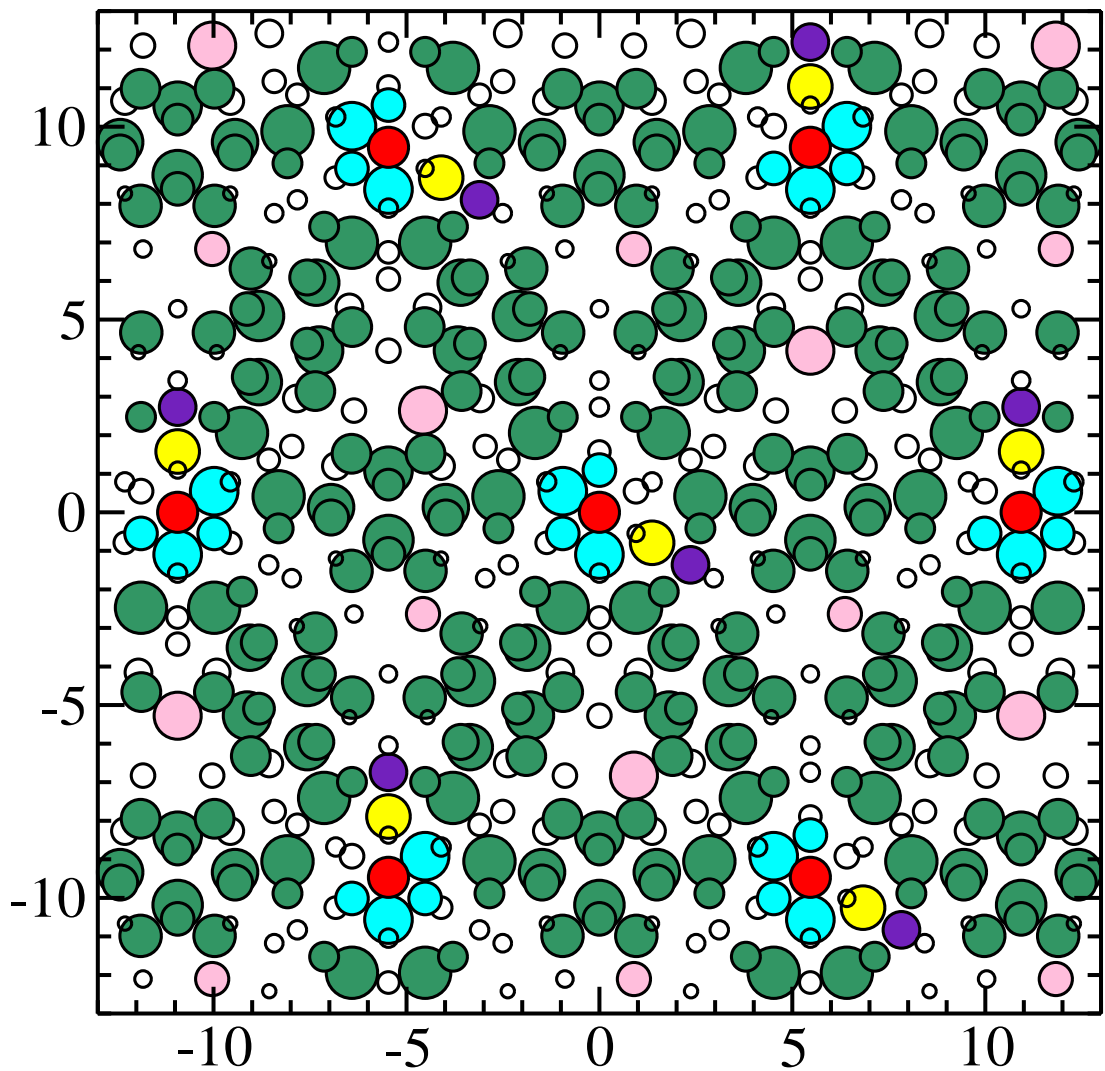}
\caption{\label{fig:opt} Optimal site occupation of a
$\beta$-rhombohedral 2x1x1 supercell viewed along the rhombohedral
(111) axis.  Color coding as in Fig.~\ref{fig:B15}.  Small empty
circles locate vacant sites.}
\end{figure}

However, within a 2x1x1 supercell of the 107 atom primitive cell,
which places independent primitive cells adjacent to each other (see
Fig.~\ref{fig:opt}), we did find superstructures which lowered the
energy significantly.  To describe these we focus on a motif near the
center of our optimal primitive cell structure. One of the two B13
vacancies is collinear with the occupied B17 and B18 sites
(e.g. B13aB17aB18a). The second B13 vacancy is adjacent to the first
(e.g. B13d).  This motif of collinear and adjacent sites can occur in
six rotated variants and six more that are reflected versions.

We examined all 22 symmetry-inequivalent arrangements of this motif
within the 214 atom supercell.  For each arrangement we optimized the
placement of the B16 atoms.  The relative arrangements are coupled
mainly by the placement of B16 atoms, and all energies lay within 2
meV/atom of each other.  Energies are given in Table~\ref{tab:scell}
for a 1x3x3 $k$-point mesh and precision set to ``Accurate''.  Several
supercell structures yielded energies lower than the optimal 107 atom
structure, indicating a preference for superlattice ordering at low
temperatures, with adjacent cells of $\beta$ resolving their partial
occupancy with differing orientations of a common motif.

The large number of nearly degenerate configurations suggests the
possibility of a phase transition from the symmetry-broken low energy
structure to a state that restores the $\beta$-rhombohedral symmetry by
sampling the full ensemble of motif orientations.  To judge the
chance of such a phase transition, we evaluate the partition function
of our 2x1x1 supercell
\begin{equation}
Z=\sum_{\alpha} \Omega_{\alpha} e^{-E_{\alpha}/k_B T}
\end{equation}
where $\alpha$ runs over all 22 symmetry-independent configurations,
$\Omega_{\alpha}$ is the multiplicity of the configuration and
$E_{\alpha}$ is the relaxed energy, optimized over placements of B16
atoms (as in Table~\ref{tab:scell} but multiplied by 214 for the
number of atoms per supercell).

\begin{figure}
\includegraphics[width=5in,angle=-90]{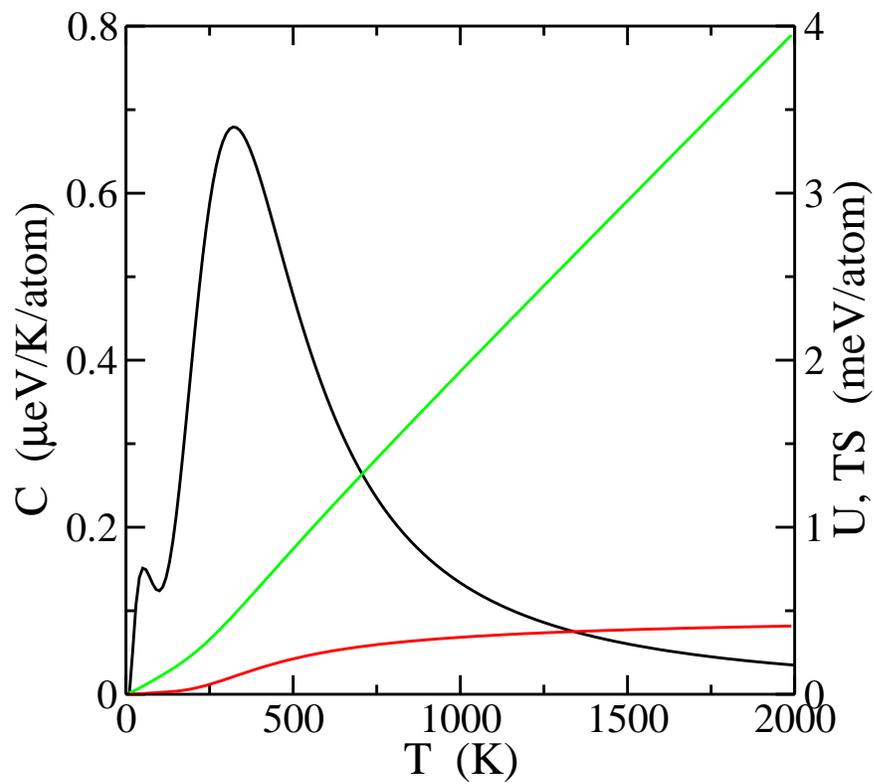}
\caption{\label{fig:thermo} Heat capacity (black), energy (red) and
entropy (green) of supercell model.  Energy is relative to the optimal
supercell structure.}
\end{figure}

Thermodynamic derivatives of $Z$ yield the internal energy $U$, the
entropy $S$ and the heat capacity $C$.  The heat capacity exhibits a
strong peak around $T=300K$.  Fig~\ref{fig:thermo} plots thermodynamic
data resulting from this model.  The strong peak around T=300K
represents the unlocking of the relative orientations of our collinear
motif (see above).  The small heat capacity peak around T=50K
represents the unlocking of the second B13 vacancies (see above) while
holding the collinear motif fixed.  The entropy available from
unlocking dominates the energy cost, substantially lowering the free
energy.  Based on this it seems likely that in an infinite system
there should be a symmetry-restoring phase transition at a moderate
temperature, driven by the entropy of suboptimal occupation patterns
of the partially occupied sites.  This transition may be difficult to
observe experimentally because atomic diffusion is slow at this
temperature.

\section{Discussion}
\label{sec:discuss}

The symmetry space group of $\beta$-rhombohedral boron is $R\bar{3}m$
(group \# 166).  To preserve rhombohedral symmetry in the primitive
cell, every Wyckoff position must be fully occupied or fully vacant.
If the favored structure occupies only a subset of the sites in one or
more Wyckoff position, the symmetry is necessarily lower.  Our optimal
vacancy-ordered structure within the primitive cell yields space group
P1 (group \#1) corresponding to no point symmetry whatsoever.

According to Landau theory~\cite{LL} changes of symmetry
occur through thermodynamic phase transitions, implying that the
$\beta$-rhombohedral structure is not the low temperature stable
phase.  Experimental observation of $\beta$ as stable indicates either
that vacancies are frozen in a disordered arrangement or that $\beta$
is stabilized by a symmetry-restoring phase transition at some low
temperature.  This conclusion holds regardless of whether we have
found the true optimal structure, since all energetically plausible
variants of $\beta$ lack full symmetry.

The transition temperature is only a crude estimate because of several
approximations.  We consider only the coupling of primitive cells.  We
consider only a small fraction of all configurations within these two
cells.  We neglect atomic vibrations which contribute strongly to the
free energy~\cite{Masago06,Setten07} and must be included in any
attempt to estimate an accurate transition temperature.  However,
while the $\alpha$ and $\beta$ phases differ strongly in vibrational
free energy, the ensemble of nearly optimal symmetry-broken $\beta$
variants most likely remains nearly degenerate in vibrational free
energy because their local environments are so similar.

Because our calculated energy differences are at the borderline of
reliability of density functional methods it would be desirable to
study these energies using more sophisticated quantum chemical
methods.  Unfortunately at present these methods cannot be reliably
applied to periodically repeated crystal structures of the necessary
complexity.  Although details of the optimal structures may change,
our primary conclusions are unlikely to be altered by higher accuracy.
The true ground state will be either symmetry-broken $\beta$ or
perhaps $\alpha$.  In either case a phase transition is necessary to
recover the experimentally observed equilibrium state above some
moderate temperature.

Even within the density functional theory limitations there remains
some uncertainty concerning the optimal structure.  Within a single
primitive cell we have only explored a fraction of all configurations
and some other unexamined structure might turn out to be prefered.
Among supercells we have only addressed the hexagonal unit cell and
the 2x1x1 supercell.  It is quite possible that a larger supercell
will be even more favorable.  In fact it could be that the true ground
state even restores rhombohedral symmetry, though in a supercell
so that it is no longer the $\beta$ structure.

In conclusion, we show that suitably resolving correlations among
partially occupied sites of $\beta$ yields an optimal structure whose
energy is lower than $\alpha$.  This structure breaks the symmetry of
$\beta$ because it uses only subsets of the fully symmetric Wyckoff
positions.  Further, we demonstrate the likelihood of superlattice
ordering.  We propose that the full symmetry of $\beta$ is restored at
moderate temperatures through a phase transition driven by the entropy
of partial site occupation.

\begin{acknowledgements}
This research was supported in part by the NSF under grant
DMR-0111198. We acknowledge useful discussions with B. Widom
\end{acknowledgements}

\bibliography{boron}

\begin{thebibliography}{33}
\bibitem[{Wyc()}]{Wyckoff}
\bibinfo{note}{A "Wyckoff position" is a group of symmetry-equivalent atomic
  sites.}
\end{thebibliography}

\newpage

\begin{table}
\begin{tabular}{|r|l|r|r|r|r|r|l|}
\hline
Name       & Pearson & Atoms & $V$   & $E-E_{\alpha}$& $\Delta R$ & Comments \\
\hline
$\alpha$-T & tP50    & 50    & 7.67  & 91.91 & 0.057 & full occupation \\
$\beta$-T  & tP196   & 192   & 7.53  & 15.13 & 0.056 & optimized occupation \\
\hline
$\beta$-R  & hR105   & 105   & 7.72  & 25.87 & 0.023 & full occupation \\
$\alpha$-R & hR12    & 12    & 7.18  &  0.00 & 0.002 & full occupation \\
\hline
$\beta$-R  & hR111   & 105   & 7.69  & 13.02 & 0.013 & B13bcdefB16a \\
$\beta$-R  & hR111   & 106   & 7.64  &  0.15 & 0.012 & B13bcdefB16bd \\
$\beta$-R  & hR141   & 107   & 7.57  & -0.86 & 0.005 & B13bcefB16bdB17aB18a \\
$\beta$-R  & hR141   & 108   & 7.56  &  1.43 & 0.015 & B13bcefB16acdB17aB18a\\
\hline
 2x1x1     & aP282   & 214   & 7.57  & -1.75 &       & optimized supercell\\
\hline
\end{tabular}
\caption{\label{tab:nrgs} Structural data including atoms per
primitive cell, volume (\AA$^3$/atom) and energy relative to
$\alpha$-R (meV/atom).  $\Delta R$ (\AA) measures the deviation of the
symmetry-averaged relaxed positions from the crystallographically
reported positions, averaged over the fully occupied Wyckoff
positions.  Comments list occupied sites using notation in
Fig.~\ref{fig:B15}. From top to bottom: tetragonal, fully occupied
rhombohedral, rhombohedral symmetry-broken structures and our
optimized supercell structure.}
\end{table}

\begin{table}
\begin{tabular}{|l||r|r||r|r||r|r||r|r|}
\hline
$\Delta E$ & 
\multicolumn{4}{c||}{$E_{\beta}-E_{\alpha}$} &
\multicolumn{4}{c|}{$E_{opt}-E_{\alpha}$} \\
\hline
$E_{cut}$ & 
\multicolumn{2}{c||}{319 eV} & \multicolumn{2}{c||}{415 eV} &
 \multicolumn{2}{c||}{319 eV} & \multicolumn{2}{c|}{415 eV} \\
\hline
Precision &
Med & Acc & Med & Acc &
Med & Acc & Med & Acc \\
\hline
\hline
$k$=1x1x1 &  -7.76 &  -7.54 &  -7.98 &  -7.97 &
        -17.27 & -17.08 & -17.04 & -17.15 \\
$k$=2x2x2 &  23.58 &  23.78 &  23.85 &  23.77 &
         -1.71 &  -1.47 &  -1.47 &  -1.68 \\
$k$=3x3x3 &  25.71 &  25.88 &  26.09 &  25.99 &
         -1.05 &  -0.86 &  -0.65 &  -0.85 \\
$k$=4x4x4 &  25.29 &  25.39 &  25.65 &  25.49 &
         -1.04 &  -0.88 &  -0.64 &  -0.87 \\
\hline
\end{tabular}
\caption{\label{tab:conv} Test of convergence as function of energy
cutoff, VASP precision setting and $k$-point mesh. Energies of hR105
($E_{\beta}$) and our optimal B13bcefB17aB18a structure ($E_{opt}$)
are compared with hR12 ($E_{\alpha}$).}
\end{table}

\begin{table}
\begin{tabular}{|l||r|r||r|r|}
\hline
$\Delta E$ & 
\multicolumn{2}{c||}{$E_{\beta}-E_{\alpha}$} &
\multicolumn{2}{c|}{$E_{opt}-E_{\alpha}$} \\
\hline
Precision &Medium & Accurate & Medium & Accurate \\
\hline
\hline
LDA   &  47.83 &  47.86 &  15.48 &  15.46\\
USPP  &  27.82 &  27.56 &  -0.25 &  -0.45\\
HARD  &  36.06 &  25.94 &  -0.59 &  -0.75\\
\hline
\end{tabular}
\caption{\label{tab:pot} Test of alternate potentials.  All calculations are
done with a 3x3x3 mesh at the default energy cutoff. USPP and HARD use
the PW91 GGA.  LDA and HARD use PAW potentials.}
\end{table}

\begin{table}
\begin{tabular}{|cccccc|cccccc||r|r|}
\hline
\multicolumn{6}{|c|}{B13 sites} & \multicolumn{6}{c||}{B16 sites} & Atoms & $E-E_{\alpha}$ \\
\hline
\hline
a&b&c&d&e&f&    & & & & & & 105 & 25.84\\
a&b&c&d&e&f&   a& & & & & & 106 & 12.25\\
a&b&c&d&e&f&    &b& &d& & & 107 & 12.50\\
\hline
 &b&c&d&e&f&    & & & & & & 104 & 27.38\\
 &b&c&d&e&f&   a& & & & & & 105 & 13.02\\
 &b&c&d&e&f&   a&b& & & & & 106 & 13.81\\
 &b&c&d&e&f&    &b& & & & & 106 & 13.10\\
\hline
 &b&c&d&e&f&    &b& &d& & & 106 &  0.15\\
 &b&c&d&e&f&    &b& & &e& & 106 &  5.21\\
 &b&c&d&e&f&    &b& & & &f& 106 &  0.87\\
\hline
\end{tabular}
\caption{\label{tab:hR111} Table of selected data for hR111. (top)
Full B13 occupation; (middle) Single B13 vacancy with single or double
occcupation of one B16 triangle; (bottom) Single occupancy of both B16
triangles.  Energy units are meV/atom.}
\end{table}

\begin{table}
\begin{tabular}{|cccccc|cccccc|c|c|c|c||r|r|}
\hline
\multicolumn{6}{|c|}{B13 sites} & \multicolumn{6}{c|}{B16 sites} & B17 & B18 & B19 & B20 & Atoms & $E-E_{\alpha}$ \\
\hline
\hline
a&b&c&d&e&f&    &b& &d& & &    a &   &   &   & 108 & 6.62 \\ 
 &b&c&d&e&f&    &b& &d& & &    a &   &   &   & 107 & 20.29 \\ 
 &b&c& &e&f&    &b& &d& & &    a &   &   &   & 106 & 0.15 \\
\hline
 &b&c& &e&f&   a& & &d& & &    a & a &   &   & 107 & -0.07 \\
 &b&c& &e&f&    &b& &d& & &    a & a &   &   & 107 & -0.86 \\
 &b&c& &e&f&    & &c&d& & &    a & a &   &   & 107 & 3.48 \\
\hline
 &b&c& &e&f&    &b& &d& & &      &   & a &   & 106 & 12.96 \\
 &b&c& &e&f&    &b& &d& & &    a & a &   & h & 108 & 0.844 \\
\hline
\end{tabular}
\caption{\label{tab:hR141} Table of selected data for hR141. (top)
B13 vacancies; (middle) B16 occupation; (bottom)
Structures occupying B19 and B20 sites.  Energy units are meV/atom.}
\end{table}

\begin{table}
\begin{tabular}{|c|c|c||c|c|}
\hline
B13(1)(2)(3) & B16(1)(2)(3) & B17B18(1)(2)(3) & Atoms & $\Delta E$\\
\hline
\hline
(bcef)(bcef)(bcef) & (bd)(bd)(bd) & (aa)(aa)(aa) & 321 & 0    \\
(bcef)(acde)(bcef) & (bd)(cf)(bd) & (aa)(bb)(aa) & 321 & 0.09 \\
(bcef)(acde)(abdf) & (bd)(cf)(ae) & (aa)(bb)(cc) & 321 & 0.31 \\
\hline
\end{tabular}
\caption{\label{tab:ucell} Hexagonal unit cell energies. Site occupancy
is given for each primitive cell, (1), (2) and (3).  $\Delta E$ is
relative to a supercell of the optimal 107 atom hR141 structure, in
units of meV/atom.}
\end{table}

\begin{table}
\begin{tabular}{|c|c|c||c|c|}
\hline
B13(1)(2) & B16(1)(2) & B17B18(1)(2) & $\Omega$ & $\Delta E$\\
\hline
\hline
(bcef)(acde) & (ad)(bd) & (aa)(bb) &  8 & -0.89 \\ 
(bcef)(acef) & (ad)(bd) & (aa)(bb) &  4 & -0.75 \\ 
(bcef)(bcdf) & (ce)(ad) & (aa)(ee) &  8 & -0.24 \\ 
(acde)(bcdf) & (bd)(ad) & (bb)(ee) &  8 & -0.11 \\ 
(bcef)(bcef) & (bd)(bd) & (aa)(aa) & 12 &  0.00 \\ 
\hline
\end{tabular}
\caption{\label{tab:scell} Selected 2x1x1 supercell energies,
including multiplicities $\Omega$.  $\Delta E$ is relative to a
doubling of the optimal 107 atom hR141 structure, in units of
meV/atom.}
\end{table}

\newpage

\newpage

\newpage

\newpage

\newpage

\newpage

\end{document}